# Unravelling Single Atom Catalysis: The Surface Science Approach

Gareth S. Parkinson

Institute of Applied Physics, TU Wien, Vienna, Austria

Email: parkinson@iap.tuwien.ac.at

This work was supported by the Austrian Science Fund START prize Y 847-N20.

Abstract

Understanding the fundamental mechanisms of single-atom catalysis (SAC) is important to design systems with improved performance and stability. This is problematic, however, because single-atom active sites are extremely difficult to characterize with existing experimental techniques. Over the last 40 years, surface science has provided the fundamental insights to understand heterogeneous catalysis, but model systems in which metal atoms are stable on well-characterized metal-oxide substrates at reaction temperatures are scarce. In this perspective, I discuss what is already known about isolated metal atoms adsorbed on model metal-oxide surfaces, and how this information can be used to understand SAC. A key issue is that, although the highly-idealised model systems studied in surface science may not be representative of a real working catalyst, they do very much resemble what can be calculated using state-of-the-art theoretical modelling. Thus, surface science offers an opportunity to rigorously benchmark the theoretical approach to modelling SAC in future. Perhaps more excitingly, several groups have developed model systems where metal adatoms remain stable at elevated temperatures. To date however, there has been no clear demonstration of catalytic activity. The perspective closes with a brief discussion of the prospect for STM experiments under realistic reaction conditions.

Main Text

The rapidly emerging field of single-atom catalysis (SAC) aims to slash the precious metal loading in heterogeneous catalysts by replacing metal nanoparticles with so-called "single-atom" active sites [1]. While there are many reports of active SAC systems, the topic remains somewhat controversial because it is very difficult to characterize a system based on single atoms, and to distinguish between these and subnano particles [2]. In practice, most groups use aberration-corrected transmission electron microscopy (TEM) to demonstrate the atomic dispersion, sometimes supplemented by XANES, which can rule out significant metal-metal bonding [1]. The activity of the catalyst is then tested, and although some mechanistic information can be drawn by in-situ techniques such as IRAS, the catalytic mechanism is proposed largely on the basis of theoretical calculations. Such calculations are based on an idealised model of the system, in which both the support structure and active site geometry are assumed. Thus, a one-to-one correspondence between the experimental and theoretical results is difficult to prove.

Traditionally, surface science has provided mechanistic information to understand heterogeneous catalysis. The idea is to strip away the complexity of a real catalyst and study well-defined single-crystal samples in a highly-controlled ultrahigh vacuum (UHV) environment. This way, the adsorption of individual reactants can be studied in detail, and an understanding of the basic interactions can be built up. The downside of the approach is that the highly idealised model system may not be as representative of the real catalyst as one would like, but on the upside, the model systems do strongly resemble what is calculated with density functional theory (DFT). Crucially, the structure of the model catalyst can be precisely determined from experiment, which ensures that accompanying theoretical calculations are realistic. In principle, one can obtain hard numbers for important parameters such as the adsorption energies (from thermal desorption experiments) and vibrational frequencies (from IRAS experiments), which can be used to benchmark the theoretical approach.

As one of the few techniques capable of true atomic resolution, scanning tunneling microscopy (STM) is potentially ideal to study SAC. In STM, an atomically sharp metallic tip is brought to within a few nm of a sample surface, and a bias from a few mV up to a few V is applied. Electrons tunnel into, or out of the sample depending on the applied bias, resulting in a tunneling current of the order nA. Since the tunneling current is exponentially dependent on the tip-sample distance, atomic-resolution images are obtained by scanning the tip over the surface and recording the tip movements necessary to maintain a constant tunneling current. There are a few important provisos, however, that must be fulfilled to perform a successful experiment. First, the support needs to be extremely flat (so that only the terminal atom of the tip interacts with the surface at any one time), which largely restricts the method to highly-oriented single crystal substrates. Second, the sample must be sufficiently conductive that a stable tunneling current can be established. The latter condition is problematic because many catalyst supports are semiconducting metal-oxides. Over the years, several strategies have been developed to circumvent the conductivity issue. Of course, some metal oxides are intrinsically conductive (e.g. $Fe_3O_4$), while others can be rendered conductive by vacuum annealing (e.g. rutile $TiO_2$). If this fails, materials can be extrinsically doped (e.g. Nb-doping of anatase $TiO_2$ or $SrTiO_3$). An alternative, and now common approach [3] is to grow the oxide support as an epitaxial thin film on a metallic substrate, which provides the requisite conductivity.

To apply the surface science approach to study single atom catalysis requires that metal adatoms are stable on a well-defined metal-oxide support. This is not straightforward because, just as in real catalysts, there is a strong thermodynamic driving force for isolated adatoms to agglomerate into metallic nanoparticles. One option is to prepare a clean, flat, conductive metal oxide surface in UHV, and then sublimate metal atoms directly onto the support at low temperature. The Freund group at the Fritz Haber Institute in Berlin have specialized in such experiments in recent years, and Figure 1 shows an example of their work on the Au/MgO(100)/Mo(100) system. Figure 1a contains an STM image acquired after Au adatoms were sublimated onto two monolayers of MgO(100) at 5 K [4], and the system subsequently exposed to CO. The Au adatoms are imaged as bright protrusions, and CO molecules adsorbed at the MgO step edges are imaged as dark depressions (see Fig 1a,b). Adsorption of CO on an Au adatom causes it to appear "fuzzy", because the CO molecule interacts with the STM tip as it sweeps by on each scan line (Fig. 1c). Particularly interesting in this work is the use of inelastic electron tunneling spectroscopy (IETS) to study the vibrational properties of adsorbed species [5]. The $d^2I/dV^2$ spectra acquired above the Au carbonyl (Fig. 1d) exhibit a symmetric peak-dip signature at +/- 50 meV due to the excitation of the frustrated CO rotation. The blue shift of this rotational mode with respect to that observed on a clean Pt surface (+/- 35 meV) is taken as evidence of the negative charge of these species, consistent with the idea that electrons tunnel from the underlying metal into the empty 6s level of the Au when the oxide is so thin. These results are remarkable as, in general, IETS is rarely performed on metal oxides due to the lack of final states in the band gap for the inelastically tunneling electrons.

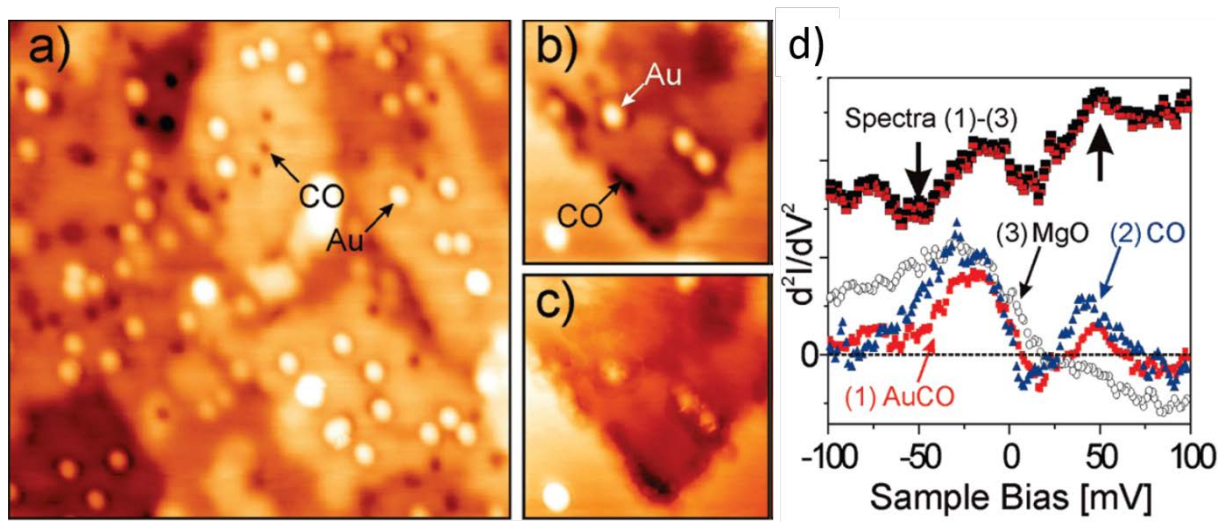

Figure 1 (a) STM image ($V_{Sample}$ = +0.1 V, $I_{tunnel}$ = 3 pA, 20×20 nm$^2$) of single Au atoms and CO molecules on MgO thin films acquired at 5 K. (b, c) High-resolution STM images of an identical surface region taken with (b) a metallic and (c) a CO-modified tip ($V_{Sample}$ = +0.1 V, $I_{tunnel}$ = 3 pA, 7.7×7.7 nm$^2$). The fuzzy appearance of the Au species in (c) occurs because the CO of the adsorbed carbonyl interacts strongly with the tip. (d) $d^2I/dV^2$ spectra taken on the bare MgO surface, a single CO molecule, and an Au-CO species (set-point: 75 mV, 10 pA). The upper curve is a difference spectrum between the Au-CO and bare MgO. The symmetric peak/dip structure at +/- 50 mV is assigned to excitation of a frustrated CO rotational mode. Figure adapted with permission from ref [4]. Copyright 2010 American Chemical Society

In recent times, the CO-stretching frequency measured in IRAS has been used to differentiate between adatoms and nanoparticles [2]. The interpretation of such frequencies is somewhat controversial [6], and an important role for surface science is to provide definitive vibrational frequencies to identify different species. The above described system, Au/MgO(100), is an excellent example of what can be achieved, and illustrates how complex the situation can be [7-8]. For example, Au adatoms can be neutral, negatively charged, and positively charged depending on the adsorption site, and both neutral and negatively charged clusters can also exist. Moreover, the adsorption of a CO molecule on a neutral Au adatom on the terrace induces sufficient charge transfer from the substrate that this species appears even more red-shifted than the negatively Au charged species! Clearly, CO is not an innocent probe of the system, and care must be taken characterizing SAC on solely the basis of IRAS frequencies.

At this juncture I would like to highlight an exciting methodological development occurring in surface science today that can directly impact the characterisation of a model single-atom catalyst. The latest generation of low-temperature scanning-probe instruments combine non-contact Atomic Force Microscopy (nc-AFM) and STM in one system. This removes the limitation regarding conductive supports, but also offers unprecedented resolution [9] and additional functionalities. For example, Onoda et al. [10] demonstrated a method to directly determine the Pauling electronegativity of adsorbed adatoms on Si(111) by measuring the maximum attractive force directly above. It is easy to imagine similar experiments for metal atoms adsorbed on metal oxide supports, with the different charge state of atoms in different adsorption sites (regular terrace / oxygen vacancy etc.) of particular interest.

While much can be learned from experiments at cryogenic temperatures, the resulting surface is probably not representative of a single-atom catalyst. In general, adatoms are probably not bound at regular lattice sites at elevated temperatures, and will quickly diffuse to occupy more strongly binding defect sites. Early surface science studies identified surface oxygen vacancies as nucleation sites for metal nanoparticles [11], and the first surface science study to report catalytically-active single atoms (already in 2000 [12]) proposed acetylene cyclotrimerization to be catalysed by negatively charged $Pd_1$ species anchored at oxygen vacancies on MgO(100). It is important to note that some materials, e.g. the iron oxides [13], do not exhibit oxygen vacancies in the bulk when reduced, instead preferring to accommodate the non-stoichiometry with bulk Fe interstitials. As such, excess Fe is a more common surface defect than an oxygen vacancy.

Since SAC really came to the fore through the work of Flytzani-Stephanopoulos [14] and Zhang [1], surface scientists have begun to investigate the stability and reactivity of $Pt_1$ on ceria and iron-oxide supports. Neyman and coworkers [15] proposed that $Pt^{2+}$ cations could be stabilized within so-called $Pt-O_4$ nanopockets. Using a combination of STM, XPS and DFT, Dvorak et al. [16] demonstrated that such nanopockets can exist at step edges on $CeO_2(111)$. Specifically, they found that $Pt^{2+}$ dominates for samples with a high step density (Fig 2g-i), whereas flat, oxygen vacancy rich samples promote the formation of metallic clusters on the terraces (Figs 2d-f). Interestingly, although the $Pt^{2+}$ was found to be thermally and chemically stable, it is apparently inactive for the dissociation of $H_2$ [17], and thus ineffective for hydrogenation reactions. A full review of the findings was recently published [18], but in general, the authors conclude that $Pt^{2+}$ cations do not interact strongly with adsorbates, and that subnano Pt nanoparticles are likely responsible for catalytic activity arising in this system.

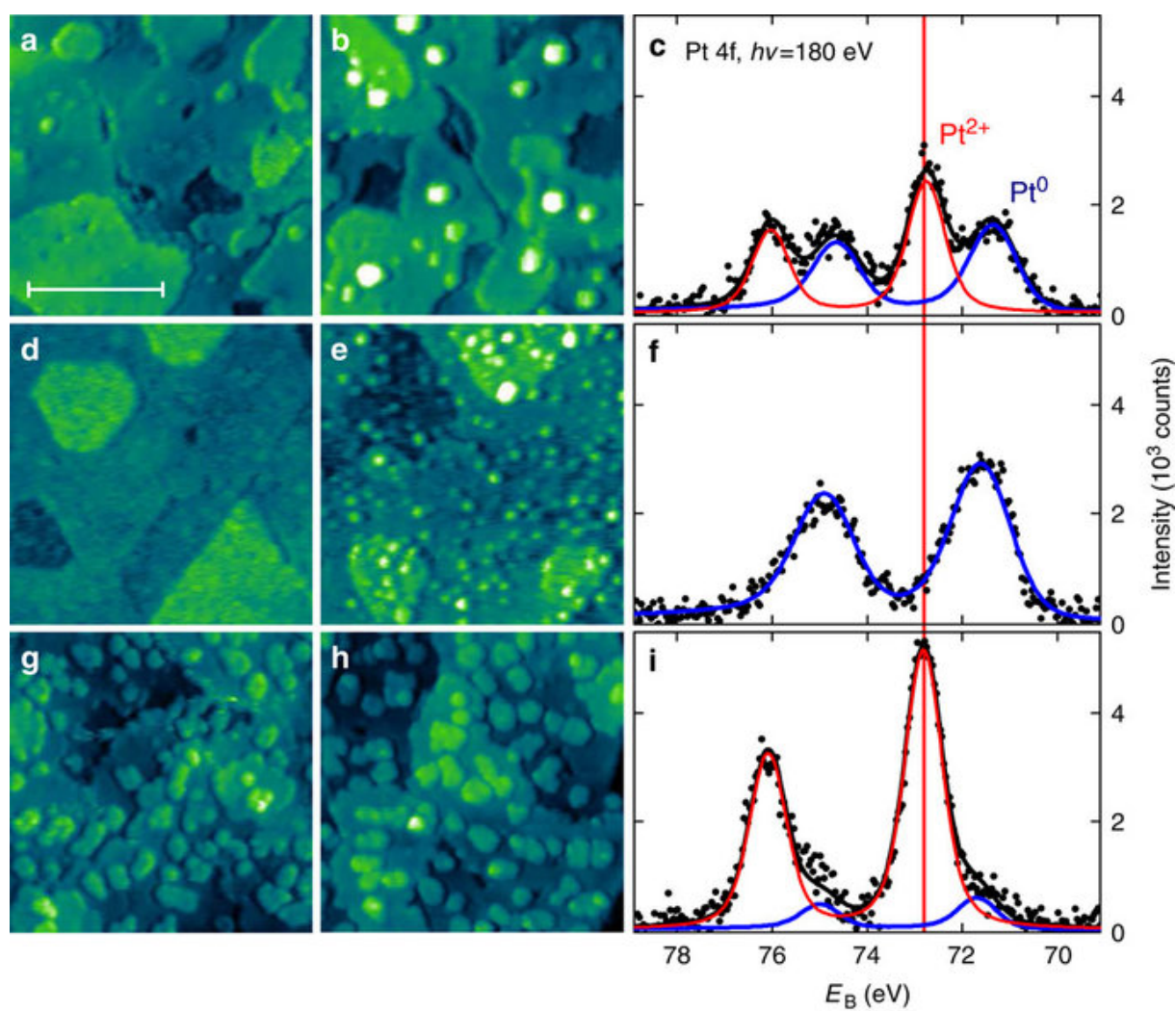

Figure 2: Characterization of a model Pt/CeO$_2$(111) catalyst by STM and XPS. The STM images (45×45 nm$^2$, V$_{sample}$ = + 2.5–3.5V, I$_{tunnel}$ = 0.25-0.75 na) show three sample types after deposition at room temperature (left column) and following annealing at 700 K in UHV (middle column). The samples are as follows: (a-b) mostly defect-free, (d-e) oxygen vacancy rich, and (g-h) high step-density. In the right column (c, f, i), corresponding Pt 4f XPS spectra are shown. Fits indicate metallic (Pt$^0$, blue line) and ionic (Pt$^{2+}$, red line) contributions to Pt 4f signal. Pt$^{2+}$ clearly dominates, and is very stable, when the ceria support has a high step density. Reproduced with permission from Ref. [13]. Copyright 2016, Nature Publishing Group.

Our group at the TU Wien has largely utilized the Fe$_3$O$_4$(001) surface as a model support because an unusual (sub)surface reconstruction [19] stabilizes arrays of adatoms (of almost any variety [13, 20]) with a periodicity of 0.84 nm to temperatures as high as 700 K. In the case of Pt [21], most adatoms are twofold coordinated to surface oxygen, as shown in Fig. 3a, and DFT+U calculations find them to be close to a neutral atom. In contrast to ceria, adatom stability is not solely due to an extremely strong interaction with the support, but also to kinetic limitations. Essentially, the Pt$_2$ dimer is unstable with respect to two Pt$_1$ adatoms on this surface, and such species rapidly decay [21]. This prevents the formation of larger clusters. STM experiments reveal that these metastable Pt atoms interact strongly with molecules from the gas phase. For example, CO adsorption lifts the Pt away from the support, as shown in Fig 3B. This species appears as a double lobed feature in STM measurements because the STM shows a time average of two symmetrically equivalent Pt-CO configurations, between which the system switches rapidly at room temperature.

One particularly advantageous aspect of the Fe$_3$O$_4$(001) model system is that diffusion is sufficiently slow at room temperature that dynamics can be followed atom-by-atom using the STM [21]. Figure 3C shows selected frames from an STM movie (in total 55 sequential STM frames recorded over the same 33x30 nm$^2$ sample area over ≈2.5 hours). The Pt$_1$ species are imaged as bright protrusions between the rows of the Fe$_3$O$_4$ support (see Fig. 3A for the configuration), and are immobile. (A second, metastable Pt$_1$* site also exists, also twofold coordinated to surface oxygen, but along the surface rows instead of across.) Pt$_1$-CO carbonyls form readily on CO exposure, appearing as the bright double-lobed feature described above. The weakened interaction with the oxide support allows the Pt$_1$-CO species to diffuse across the surface, and when they meet, stable Pt$_2$(CO)$_2$ dimers are formed. There are instances, such as that highlighted in the figure, where larger clusters form through the same mechanism. Interestingly, the Pt$_2$(CO)$_2$ dimers break apart into two Pt$_1$ species when the CO desorbs at ≈500 K, resulting in the recovery of the Pt adatom phase. Ultimately however, we find that the Pt/Fe$_3$O$_4$(001) system is active for CO and H$_2$ oxidation through an MvK-type O$_{lattice}$ abstraction at ≈ 520 K [22] only when subnano Pt clusters are present. Pt adatoms do not participate because they sinter too rapidly upon exposure to CO.

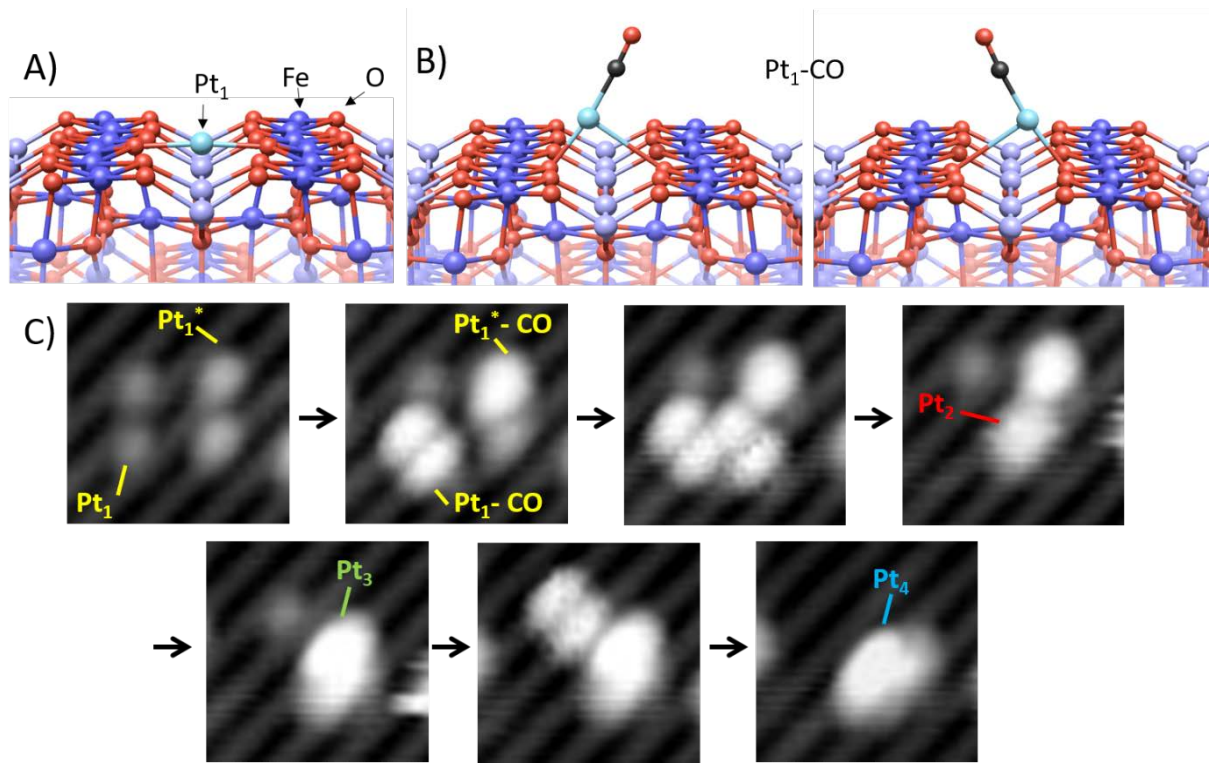

Figure 3. CO-induced Pt diffusion and coalescence on $Fe_3O_4(001)$-$(\sqrt{2}\times\sqrt{2})R45°$. (A) The energetically preferred $Pt_1$ geometry is two-fold coordinated to surface oxygen atoms across the rows. A metastable geometry ($Pt1^*$) in which Pt is bound to two O atoms along the row is not shown. (B) Adsorption of a CO molecule lifts the Pt atom from the surface, resulting in a Pt-CO species that switches rapidly between two symmetrically equivalent configurations at room temperature. This produces a double-lobed appearance in STM. (C) STM image sequence acquired during exposure to $2\times10^{-10}$ mbar CO showing the formation of a Pt tetramer: CO adsorption on $Pt_1$ and $Pt_1^*$, mobility, and coalescence are observed atom by atom. Figure adapted from ref. [21] with permission from the authors.

As a final example, we consider the recent work of Zhou et al. [23], who successfully combined atomically-resolved STM with temperature programmed desorption (TPD) to study the reactivity of Pt atoms and subnano particles supported by a monolayer of CuO grown on Cu(110) (see Fig. 4). The authors demonstrate that subnano clusters catalyse CO oxidation activity via a Mars-van Krevelen (MvK) mechanism, and that Pt adatoms are inactive because CO desorbs at ≈ 300 K from the carbonyl, long before there is sufficient thermal energy available in the system to extract O from the lattice. Such weak CO binding on the Pt adatom is in stark contrast to the behaviour on the $Fe_3O_4(001)$ support described above. Unfortunately, it is difficult to ascertain the reason at present because no spectroscopy or DFT calculations were performed with which one could ascertain the oxidation state of the $Pt_1$ species.

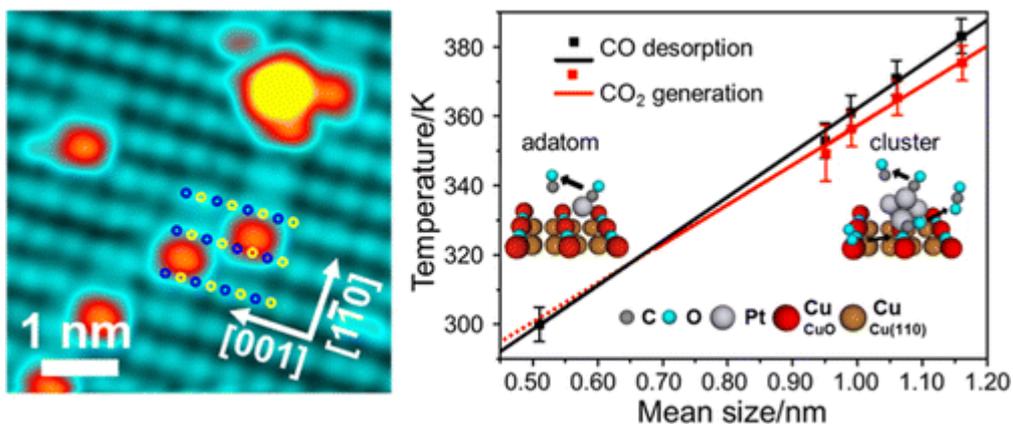

Figure 4: STM image ($V_{sample}$ = +0.3 V, $I_{tunnel}$ = 30 pA) of Pt single atoms and nanoclusters on CuO monolayer prepared by deposition of 0.1 ML Pt at RT. Blue and yellow circles represent the Cu and O ions of the CuO support, respectively. (Right) Plots of the CO desorption and $CO_2$ generation temperatures versus mean size of the Pt nanoclusters. In the case of a single Pt atom, CO desorbs before there is sufficient thermal energy to allow a MvK reaction with the underlying substrate. Figure adapted with permission from ref [23]. Copyright 2016 American Chemical Society

In the three examples cited above there appears to be precious little evidence to support catalytic activity of oxide supported Pt atoms. Undercoordinated, neutral Pt adatoms interact too strongly with CO, leading to rapid sintering. On the other hand, highly coordinated $Pt^{2+}$ cations do not seem to interact strongly enough with adsorbates to be catalytically active. It must be noted however, that all of the above-mentioned work was conducted under UHV conditions, and it is certainly possible that a significant pressure gap exists. Although CO desorbs from $Pt_1$ on CuO at 300 K in a TPD experiment, at high pressures the relatively weak interaction could still be sufficient to trap some CO molecules near the surface long enough for a MvK process to occur. To investigate SAC at such conditions one needs to perform STM at high pressures and temperatures. Both situations are, in principle, relatively straightforward to achieve, and commercial setups have been available for more than a decade. Nevertheless, experiments are difficult because STM is a comparatively slow technique, and not suited to study systems with rapid diffusion. Similarly, STM at high pressure is possible, but the hard fought certainty and control offered by UHV conditions is sacrificed, and assigning the different protrusions becomes impossible. There have been notable successes [24], particularly revealing relatively large scale changes to surface morphology that can occur under reaction conditions. Unravelling the mechanisms of a single atom catalyst under reaction conditions requires a level of resolution that is yet to be demonstrated. At present, several groups around the world are currently working to make this reality however, and we eagerly await their results.

It seems almost premature to discuss future directions of SAC in surface science when so much basic work remains to be done. In the dream scenario, lessons learned from surface science will be used to directly improve catalytic systems, or maybe even allow the design of a new one from scratch. This ideal is exemplified in the fruitful collaboration between the Sykes and Flytzani-Stephanopoulos groups at Tufts University. In their work on the related concept of single atom alloys (SAA) [25], a concept is typically demonstrated in UHV experiments, and the results used to design an analogous nanoparticle system, which is then tested for catalytic activity [26]. These articles benefit hugely from the dual-pronged approach, and in my opinion, this strategy should be applied where possible as we attempt to understand the reactivity of single-atom catalysts based on oxide supports.

Acknowledgements

GSP acknowledges funding from the Austrian Science Fund START prize Y 847-N20, and thanks Prof. Ulrike Diebold (TU Wien) for critically reading the manuscript.

# Unravelling Single Atom Catalysis: The Surface Science Approach

Gareth S. Parkinson

*Institute of Applied Physics, TU Wien, Vienna, Austria*

Email: parkinson@iap.tuwien.ac.at

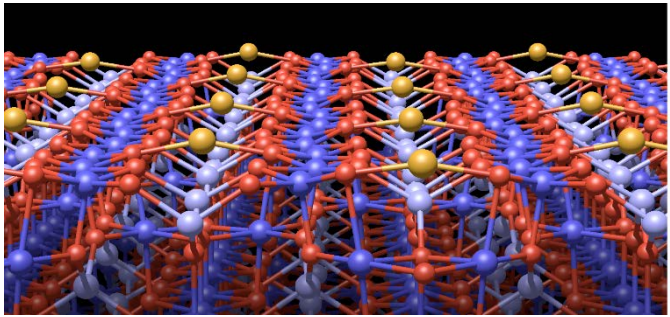

This perspective discusses how studies of idealised model systems can shed light on the fundamental mechanisms of single-atom catalysis. The image shows Au adatoms supported by $Fe_3O_4$(001).